\documentclass[aps,prd,preprint,groupedaddress,showpacs]{revtex4-1}

\usepackage{graphicx}
\usepackage{color}

\newcommand{\bfv}{\mbox{\boldmath $v$}}
\newcommand{\bfx}{\mbox{\boldmath $x$}}
\newcommand{\bfB}{\mbox{\boldmath $B$}}

\begin{document}


\title{Updated constraint on a primordial magnetic field during big bang nucleosynthesis and a formulation of field effects}


\author{Masahiro Kawasaki$^{1,2}$ and Motohiko Kusakabe$^{1}$
}
\email{kusakabe@icrr.u-tokyo.ac.jp}
\affiliation{
$^1$Institute for Cosmic Ray Research, University of Tokyo, Kashiwa,
Chiba 277-8582, Japan\\
$^2$Kavli Institute for the Physics and Mathematics of the Universe, University of Tokyo, Kashiwa, Chiba 277-8582, Japan}


\date{\today}

\begin{abstract}
A new upper limit on the amplitude of primordial magnetic field (PMF) is derived by a comparison between a calculation of elemental abundances in big bang nucleosynthesis (BBN) model and the latest observational constraints on the abundances.  Updated nuclear reaction rates are adopted in the calculation.  Effects of PMF on the abundances are consistently taken into account in the numerical calculation with the precise formulation of changes in physical variables.  We find that abundances of $^3$He and $^6$Li increase while that of $^7$Li decreases when the PMF amplitude increases, in the case of the baryon-to-photon ratio determined from the measurement of cosmic microwave background radiation.  We derive a constraint on the present amplitude of PMF, i.e., $B(0)<1.5~\mu$G [corresponding to the amplitude less than $2.0\times 10^{11}$~G at BBN temperature of $T=10^9$~K] based on the rigorous calculation.
\end{abstract}


\maketitle

\section{Introduction}\label{sec1}

Primordial nucleosynthesis, or big bang nucleosynthesis (BBN), has been assumed~\cite{Gamow:1946eb} to occur through complicated nonequilibrium processes.  It involves many reactions including radiative neutron capture reactions~\cite{Gamow:1946eb,Alpher:1948ve,Alpher1948b} and weak interactions converting protons and neutrons to each others~\cite{Hayashi1950} as well as relativistic quantum statistics~\cite{Alpher1953}.  In BBN, only D, $^3$He, $^4$He and $^7$Li can be produced in significant amounts~\cite{Wagoner:1966pv}, and yields of heavier elements are generally expected to be small~\cite{Hayashi1956,Sato1967}.  The BBN model predicts the relic of a dense hot radiation~\cite{Alpher1948b,Alpher1949} to be observed as cosmic microwave background radiation (CMBR) today~\cite{Penzias:1965wn}.

The BBN has been studied over a long period, and the theory is now precisely structured (e.g.,~\cite{Wagoner:1966pv,Fowler1967,Wagoner:1972jh,Olive:1980bu,Dicus1982,Audouze:1985be,Boesgaard:1985km,Olive:1989xf,Krauss1990,Smith:1992yy,Malaney:1993ah,Copi:1994ev,Krauss:1994rx,Sarkar:1995dd,Olive:1995kz,Hata:1995tt,Fields:1996yw,Schramm:1997vs,Esposito:1999sz,Burles:1999zt,Tytler2000,VangioniFlam:2000xs,Burles:2000zk,Cyburt:2001pp,Coc:2002tr,Coc:2003ce,Iocco:2008va,Mangano:2011ar,Coc:2011az}).  The simplest, standard BBN (SBBN) model is characterized by one parameter, i.e, baryon-to-photon number ratio $\eta$ with the fixed number of light neutrino species of $N=3$.  The $\eta$ value is constrained precisely with data of the Wilkinson Microwave Anisotropy Probe (WMAP) \cite{Spergel:2003cb,Spergel:2006hy,Larson:2010gs}.  The SBBN model prediction of light element abundances for the WMAP $\eta$ value is rather consistent with primordial abundances inferred from observations.  There is, however, a discrepancy between the predicted and observed primordial abundances of $^7$Li.  The SBBN predicts a $^7$Li abundance which is a factor of $2.4-4.3$ times higher \cite{Cyburt:2008kw} than the observationally deduced abundance.  Possible solutions to this discrepancy have been proposed (e.g.~\cite{Kawasaki:2010yh} and references therein).

O'Connel and Matese~\cite{OConnell1969} have estimated the neutron $\beta$-decay rate in the presence of a strong magnetic field, and suggested that an increase in the rate due to primordial magnetic fields (PMFs) decreases $^4$He abundance.  Greenstein~\cite{Greenstein1969} subsequently suggested that the energy of PMFs enhances the expansion rate of the universe, and it tends to increase the $^4$He abundance rather than decrease it as suggested in Ref.~\cite{OConnell1969}.  Matese \& O'Connel~\cite{Matese1970} then performed a detailed investigation on the PMF effects on BBN, and concluded that the effect through the expansion rate is predominant over that through rates of weak reactions.

Three groups have investigated effects of PMF on BBN~\cite{Cheng:1993kz,Grasso:1994ph,Kernan:1995bz,Grasso:1996kk,Cheng:1996yi,Kernan:1996ab}, and have the common conclusion that the effect through the cosmic expansion rate contributed from an enhanced energy density~\cite{Greenstein1969} is the most important~\cite{Kernan:1995bz,Grasso:1996kk,Cheng:1996yi}.  The effect of energy density of PMF can be considered in analogy with that of an effective neutrino number in BBN epoch~\cite{Kernan:1995bz}.  Grasso \& Rubinstein~\cite{Grasso:1996kk} have additionally shown that a change in the quantum statistics of electron and positron by the PMF affects BBN.  Constraints on PMF depend on other parameters than the amplitude of PMF.  Suh \& Mathews~\cite{Suh:1999va} have studied sensitivity of limits on PMF to the neutrino degeneracy.  See section 3 in Ref.~\cite{Grasso:2000wj} for a review of this topic.

The latest BBN constraint on the magnetic field~\cite{Cheng:1996yi,Grasso:2000wj} has been vary old.  It was based on assumptions of an old baryon-to-photon ratio $\eta=2.8\times 10^{-10}$ and the upper limit on $^4$He mass fraction of $Y{\rm p}\leq 0.245$.  The values are updated to be $\eta=6.2\times 10^{-10}$ for $\Lambda$CDM+SZ+lens model~\cite{Larson:2010gs}, and $Y=0.2561\pm 0.0108$~\cite{Aver:2010wq}.  In this paper we perform a network calculation of BBN taking account of effects of PMFs, and show a latest constraint on PMF through effects on elemental abundances.  In addition, formulae necessary for precise numerical calculations are provided.  This study improves the following points over previous works: 1) updates on nuclear reaction rates, the neutron lifetime, and observational constraints on primordial abundances, 2) a precise treatment of electron chemical potential in abundance calculation and an estimation for initial value of electron chemical potential, 3) a precise calculation of temperature evolution as a function of time or cosmic scale factor, and 4) a caution that an effect of magnetic field on nuclear reaction rates is weak.

In Sec.~\ref{sec2} we describe the model of SBBN code with a recent update on nuclear reaction rates and the neutron lifetime, and also how to include magnetic fields effects on BBN in precise numerical studies.  In Sec.~\ref{sec3} we show results of calculations of BBN in the presence of variable amplitudes of PMF.  In Sec.~\ref{sec4} we discuss constraints on PMFs.  In Sec.~\ref{sec5} we summarize this study.  In Appendix \ref{appendix} we describe formulae necessary for BBN network calculations including effects of PMFs.  In Appendix \ref{appendix_2} an effect of PMFs on nuclear reaction rates is studied, and it is shown to be negligible.

\section{Model}\label{sec2}

\subsection{standard BBN}\label{sec21}

We use a BBN code~\cite{Kawano1992,Smith:1992yy} for reaction network calculations. The Sarkar's correction is adopted for $^4$He abundance~\cite{Sarkar:1995dd}.  Rates and their uncertainties of reactions for light nuclei ($A \le  10$) are updated with recommendations of the JINA REACLIB Database V1.0 \cite{Cyburt2010}.  We derive 95 \% confidence regions of elemental abundances assuming uncertainties in rates of the 12 important reactions~\cite{Smith:1992yy}.  The rates are assumed to be given by the Gaussian distribution, and 1000 runs are performed for each eta value.  The reactions and the references for adopted rates are listed in Table~\ref{tab1}.

\begin{table}[!t]
\caption{\label{tab1} Adopted reaction rates}
\begin{ruledtabular}
\begin{tabular}{cccc}
ID\footnote{Reaction number in the Kawano's code \cite{Kawano1992}.} & reaction & reference \\ 
\hline
1  & $n$(,$e^- \bar{\nu_e}$)$^1$H    & \cite{Serebrov:2004zf} and \cite{Nakamura:2010zzi} \\
12 & $^1$H($n$,$\gamma$)$^2$H        & \cite{Ando:2005cz} \\
16 & $^3$He($n$,$p$)$^3$H            & \cite{Descouvemont2004} \\
17 & $^7$Be($n$,$p$)$^7$Li           & \cite{Descouvemont2004} \\
20 & $^2$H($p$,$\gamma$)$^3$He       & \cite{Descouvemont2004} \\
24 & $^7$Li($p$,$\alpha$)$^4$He      & \cite{Descouvemont2004} \\
26 & $^3$H($\alpha$,$\gamma$)$^7$Li  & \cite{Descouvemont2004} \\
27 & $^3$He($\alpha$,$\gamma$)$^7$Be & \cite{Cyburt:2008up} \\
28 & $^2$H($d$,$n$)$^3$He            & \cite{Descouvemont2004} \\
29 & $^2$H($d$,$p$)$^3$H             & \cite{Descouvemont2004} \\
30 & $^3$H($d$,$n$)$^4$He            & \cite{Descouvemont2004} \\
31 & $^3$He($d$,$p$)$^4$He           & \cite{Descouvemont2004} \\
\end{tabular}
\end{ruledtabular}
\end{table}

We adopt two values of neutron lifetime.  One is $878.5 \pm 0.7_{\rm stat} \pm 0.3_{\rm sys}$~s from Ref.~\cite{Serebrov:2010sg} based on improvements \cite{Serebrov:2004zf} in the measurement.  This relatively short lifetime better satisfies the unitarity test of the Cabibbo-Kobayashi-Maskawa matrix \cite{Serebrov:2004zf}, and it can improve the agreement between observed primordial abundances and  BBN predictions~\cite{Mathews:2004kc,Cheoun:2011yn}.  Another value is $885.7 \pm 0.8$~s from the old recommendation by the Particle Data Group \cite{Nakamura:2010zzi}.  As of April 2012, the Particle Data Group presents a new average neutron lifetime of $881.5 \pm 1.5$~s which is sandwiched between the adopted lifetimes. \footnote{This value is close to an average lifetime of $880.0 \pm 0.9$~s estimated~\cite{Serebrov:2011re} by including recent yet unpublished reports of experiments with ultracold neutrons (see their references)}.

We adopt constraints on primordial abundances as follows:

A deuterium abundance in a damped Lyman alpha system of QSO SDSS J1419+0829 was measured precisely than any other QSO absorption systems~\cite{Pettini:2012ph}.  We adopt both of a mean value of ten QSO absorption line systems including J1419+0829, and the abundance of J1419+0829 itself, i.e., log(D/H)=$-4.58\pm 0.02$ and log(D/H)=$-4.596\pm 0.009$, respectively.  We take $2\sigma$ uncertainties, i.e., 
\begin{eqnarray}
2.40\times10^{-5}< {\rm D}/{\rm H}< 2.88\times10^{-5}~~~~~{\rm (mean)},\nonumber\\
2.43\times10^{-5}< {\rm D}/{\rm H}< 2.64\times10^{-5}~~~~~{\rm (best)}.
 \label{eq25}
\end{eqnarray}

$^3$He abundances are measured in Galactic HII regions through the 8.665~GHz
hyperfine transition of $^3$He$^+$, i.e., $^3$He/H=$(1.9\pm 0.6)\times
10^{-5}$~\cite{Bania:2002yj}.  Although
the constraint is rather weak considering
its uncertainty, we take a $2\sigma$ upper limit from abundances
in Galactic HII region, i.e.,
\begin{equation}
^3{\rm He}/{\rm H}< 3.1\times 10^{-5}.
 \label{eq26}
\end{equation}

For the primordial helium abundance we adopt two different constraints,
i.e, $Y=0.2565\pm 0.0051$~\cite{Izotov:2010ca} and $Y=0.2561\pm
0.0108$~\cite{Aver:2010wq} both from observations of metal-poor extragalactic
HII regions.  We take $2\sigma$ limits of 
\begin{eqnarray}
0.2463 &< Y <& 0.2667~~~~~{\rm (IT10)},\nonumber\\
0.2345 &< Y <& 0.2777~~~~~{\rm (AOS10)}.
\label{eq28}
\end{eqnarray}

As a guide, observed lithium abundances follow although they are not used as constraints.  

Primordial $^7$Li abundance is inferred from spectroscopic observations of metal-poor halo stars (MPHSs).  We adopt log($^7$Li/H)$=-12+(2.199\pm 0.086)$ (95\% confidence limits) derived in a 3D nonlocal thermal equilibrium model~\cite{Sbordone2010}.
This estimation corresponds to the $2\sigma$ range of
\begin{equation}
1.06\times 10^{-10} < {\rm ^7Li/H} < 2.35\times 10^{-10}.
\label{eq29}
\end{equation}

Observations of MPHSs suggest a presence of $^6$Li nuclei in some of the stars.  The most probable detection of $^6$Li for G020-024 indicates $^6$Li/$^7$Li=$0.052\pm 0.017$ \cite{2010IAUS..265...23S}.  We use the $2\sigma$ upper limit and $\log(^7{\rm Li}/{\rm H})=-12+2.18$ for the same star \cite{Asplund:2005yt}, and derive
\begin{equation}
^6{\rm Li/H} < 1.3\times 10^{-11}.
\label{eq30}
\end{equation}

\begin{figure}
\begin{center}
\includegraphics[width=8.0cm,clip]{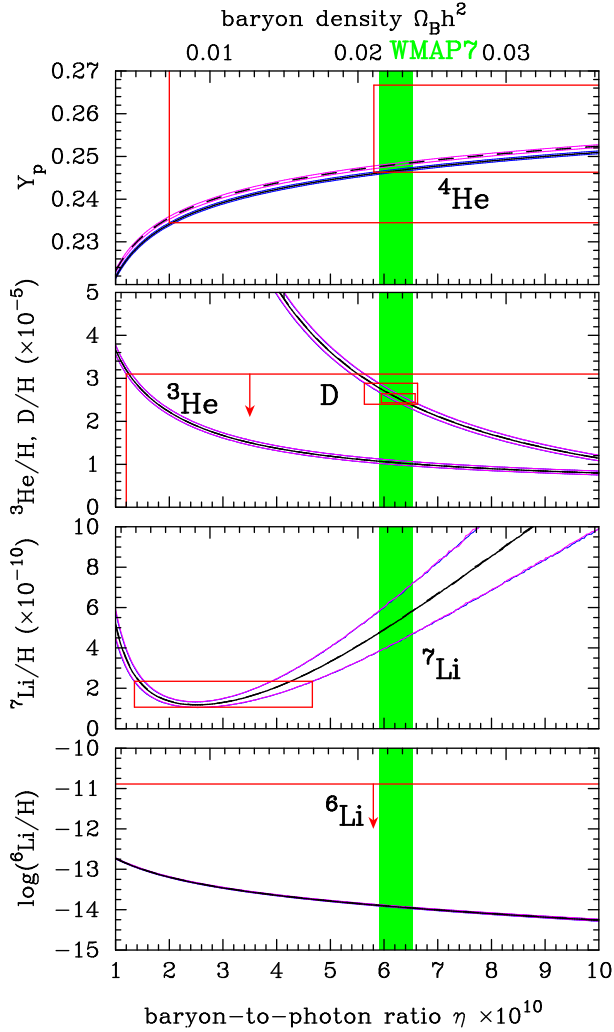}
\caption{Abundances of $^4$He (mass fraction), D, $^3$He,
 $^7$Li and $^6$Li (by number relative to H) as a function of the
 baryon-to-photon ratio $\eta$ or the baryon energy density $\Omega_B
 h^2$.  The solid and dashed curves are the results for neutron lifetimes of $878.5\pm 0.8$ s~\cite{Serebrov:2004zf} and $885.7\pm0.8$ s~\cite{Nakamura:2010zzi}, respectively.  Thin solid curves show 95 \% ranges determined from uncertainties in nuclear reaction rates. The boxes represent
 the adopted observational abundances from Refs.~\cite{Izotov:2010ca,Aver:2010wq} for
 $^4$He,~\cite{Pettini:2012ph} for D,~\cite{Bania:2002yj} for $^3$He,~\cite{Sbordone2010} for $^7$Li,
 and~\cite{2010IAUS..265...23S,Asplund:2005yt} for $^6$Li, respectively.  The vertical
 stripe represents the 2~$\sigma$~$\Omega_B h^2$ limits provided by
 WMAP~\cite{Larson:2010gs}, i.e., $\Omega_B h^2=0.02258^{+0.00114}_{-0.00112}$ or $\eta=(6.225^{+0.314}_{-0.309})\times 10^{-10}$. \label{fig1}}
\end{center}
\end{figure}

Figure \ref{fig1} shows abundances of $^4$He ($Y_{\rm p}$; mass fraction), D, $^3$He, $^7$Li and $^6$Li ($A/$H; by number relative to H) as a function of the baryon-to-photon ratio $\eta$ or the baryon energy density $\Omega_B h^2$ of the universe.  The solid and dashed curves are the results for neutron lifetimes of $878.5\pm 0.8$ s~\cite{Serebrov:2004zf} and $885.7\pm0.8$ s~\cite{Nakamura:2010zzi}, respectively.  Thin solid curves show 95 \% ranges determined from uncertainties in nuclear reaction rates. The boxes represent the adopted abundance constraints.  The vertical stripe represents the 2~$\sigma$~$\Omega_B h^2$ limits provided by WMAP~\cite{Larson:2010gs}.  This corresponds to $\Omega_B h^2=0.02258^{+0.00114}_{-0.00112}$ or $\eta=(6.225^{+0.314}_{-0.309})\times 10^{-10}$.


\subsection{effects of magnetic field}

We include effects of PMF through the magnetic energy density (Appendix \ref{appendix1}), thermodynamic variables of electron and positron and their time evolutions (Appendixes \ref{appendix2}, \ref{appendix3}).  Estimations for initial values of electron chemical potential are changed from those in the case of no magnetic field (Appendix \ref{appendix4}).  Equations to solve are similar to those in Ref.~\cite{Kernan:1995bz}.  However, equations which are solved in the consistent numerical calculation (see Appendix \ref{appendix}) are more complicated.

Final $\eta$ values are different for different initial $B$ values with a fixed initial $\eta$ value.  The final $\eta$ value, instead of the initial value, should then be fixed to the value determined from WMAP~\cite{Larson:2010gs} as pointed out but not done in deriving the limit on $B$ [eq. (4)] in Ref.~\cite{Grasso:1996kk}.  In this study we fixed the final $\eta$ value.  The effect of the magnetic field on weak reaction rates has been long since found to be negligible~\cite{Grasso:1996kk,Kernan:1995bz}.  It is, therefore, not included in this calculation.

\section{Result}\label{sec3}

\begin{figure}
\begin{center}
\includegraphics[width=8.0cm,clip]{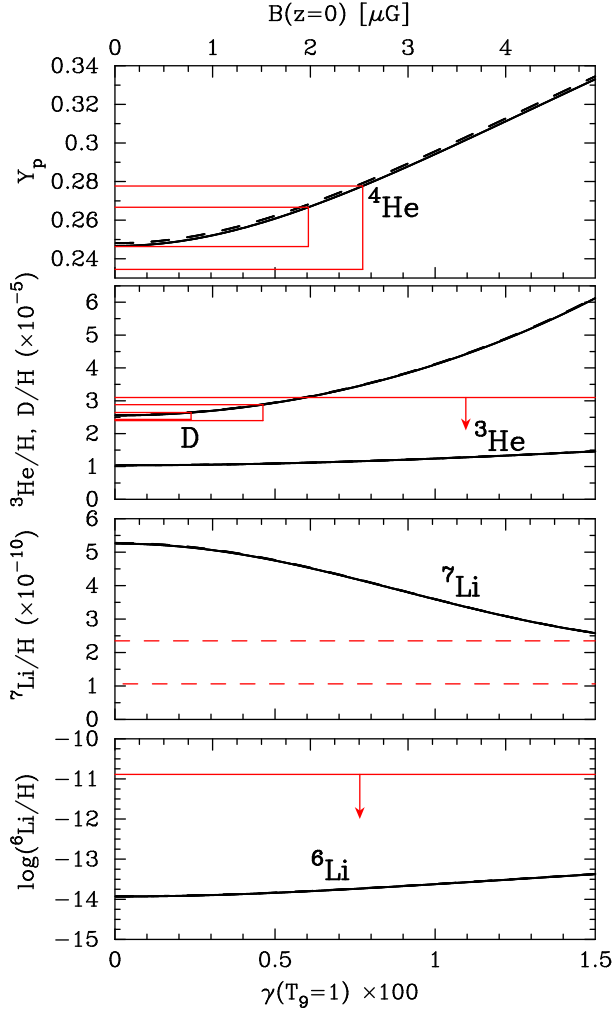}
\caption{Abundances of $^4$He (mass fraction), D, $^3$He,
 $^7$Li and $^6$Li (by number relative to H) as a function of the
 magnetic field amplitude in units of the critical value, i.e., $\gamma=B/B_{\rm C}$, or absolute value at the present epoch of $z=0$.  The solid and dashed curves are the results for neutron lifetimes of $878.5\pm 0.8$ s~\cite{Serebrov:2004zf} and $885.7\pm0.8$ s~\cite{Nakamura:2010zzi}, respectively.  The boxes represent
 adopted abundance constraints which are the same as in Fig.~\ref{fig1}. The final value of baryon-to-photon ratio is fixed to be $\eta=6.2\times 10^{-10}$~\cite{Larson:2010gs}. \label{fig2}}
\end{center}
\end{figure}

Figure \ref{fig2} shows light element abundances as a function of the
 magnetic field amplitude in units of the critical value, i.e., $\gamma=B/B_{\rm C}$ at temperature $T=10^9$~K, or absolute value at the present epoch of $z=0$, i.e., $B(0)$.  $B_{\rm C}=4.41\times 10^{13}$ G is the critical magnetic field (above which quantized magnetic levels appear~\cite{Grasso:2000wj}).  The two parameters are related by $\gamma(T=10^9$ K$)=3.05[B(0)/$mG].  The solid and dashed curves correspond to results for two different neutron lifetimes, and the boxes represent adopted abundance constraints (see Sec.~\ref{sec21}).


\begin{figure}
\begin{center}
\includegraphics[width=8.0cm,clip]{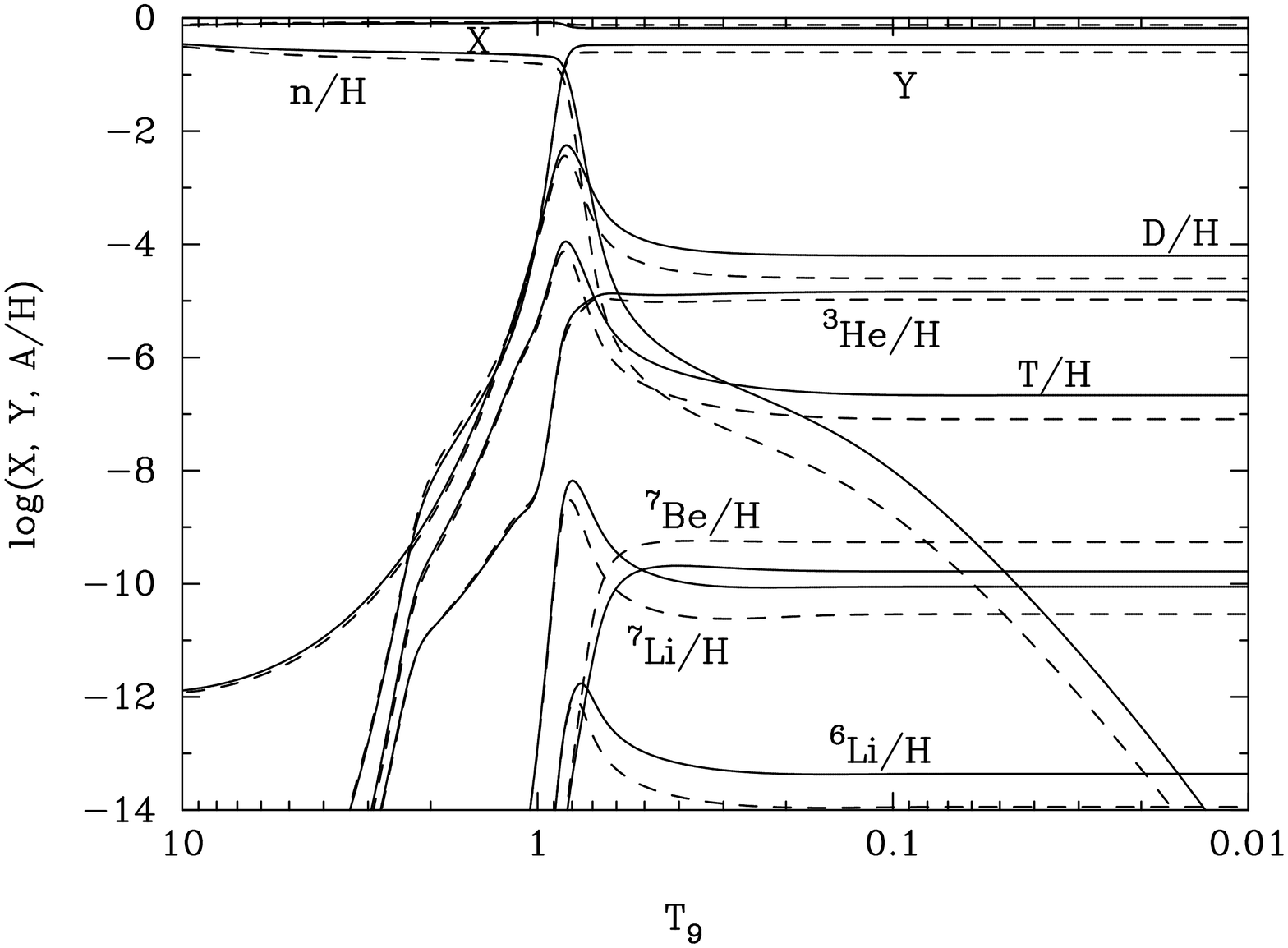}
\caption{Abundances of H and $^4$He, i.e., $X$ and $Y$, respectively, (mass fraction), and other nuclides (by number relative to H) as a function of the temperature $T_9\equiv T/(10^9~{\rm K})$.  Solid lines correspond to the case of a magnetic field of $B(z=0)=5~\mu$G, while dashed lines correspond to standard big bang nucleosynthesis model.  In both cases, final values of baryon-to-photon ratios are $\eta=6.2\times 10^{-10}$~\cite{Larson:2010gs}. \label{fig3}}
\end{center}
\end{figure}

Figure \ref{fig3} shows time evolutions of light element abundances as a function of the temperature $T_9\equiv T/(10^9~{\rm K})$.  Solid lines correspond to the case of a magnetic field of $B(z=0)=5~\mu$G, while dashed lines correspond to SBBN.


The primordial abundance of $^4$He increases when the amplitude of PMF increases.  The cosmic expansion rate is larger because of the energy of the PMF, so that the neutron abundance after the earlier freeze-out of weak reactions is higher.  The time interval between the freeze-out and the $^4$He synthesis is also shorter because of faster cosmic expansion.  Neutron abundances are larger than in SBBN for the above two reasons.  Those neutrons are processed to form $^4$He nuclei.  This is the reason of the trend of $^4$He abundance vs. $B$ value.

Because of the earlier freeze-out of the reaction $^1$H($n,\gamma$)$^2$H at the $^4$He synthesis, the relic neutron abundance is higher than in SBBN.  This higher neutron abundance affects abundances of other light nuclei complicatedly.  The abundance of D which is produced via $^1$H($n,\gamma$)$^2$H is somewhat higher when $B$ is higher.  $^3$H is produced via $^2$H($d,p$)$^3$H and destroyed via $^3$H($d,n$)$^4$He.  The enhanced D abundance simply leads to a higher $^3$H abundance by a higher production rate.  $^3$He, on the other hand, is produced via $^2$H($d,n$)$^3$He and destroyed via $^3$He($n,p$)$^3$H.  The somewhat higher D abundance leads to higher production rate while the higher neutron abundance leads to higher destruction rate.  Resultingly, the $^3$He abundance is slightly higher than in SBBN.

$^6$Li is produced via $^4$He($d,\gamma$)$^6$Li and destroyed via $^6$Li$(p,\alpha$)$^3$He, and $^7$Li is produced via $^4$He($t,\gamma$)$^7$Li and destroyed via $^7$Li($p,\alpha$)$^4$He.  The abundances of both nuclides are then higher since those of D and T are higher at higher $B$ values.  $^7$Be is produced via $^4$He($^3$He$,\gamma$)$^7$Be and destroyed via $^7$Be$(n,p$)$^7$Li.  Slightly higher abundance of $^3$He and rather higher abundance of neutron results in net reduction of $^7$Be abundance with respect to that of SBBN.

The final $^7$Li abundance is the sum of those for $^7$Li and $^7$Be.  $^7$Be nuclei convert to $^7$Li nuclei by an electron capture process.  Since the abundance of $^7$Be is larger than that of $^7$Li in SBBN, an existence of PMF reduces the final $^7$Li abundance.

Shapes of curves in Fig. \ref{fig2} are explained as above.  Increases in abundances of $^4$He and (D+$^3$He) have been obtained in the previous investigation~\cite{Grasso:1996kk}, and are consistent with our result.

The following constraints are derived from Fig.~\ref{fig2}.
\begin{eqnarray}
B(0) < 2.0~\mu{\rm G}~~~& {\rm or}~\gamma(T_9=1) < 6.0\times 10^{-3}~~~& (^4{\rm He};~{\rm IT10}),\nonumber\\
B(0) < 2.5~\mu{\rm G}~~~& {\rm or}~\gamma(T_9=1) < 7.7\times 10^{-3}~~~& (^4{\rm He};~{\rm AOS10}),
\end{eqnarray}

\begin{eqnarray}
B(0) < 1.5~\mu{\rm G}~~~& {\rm or}~\gamma(T_9=1) < 4.6\times 10^{-3}~~~& ({\rm D};~{\rm mean}),\nonumber\\
B(0) < 0.78~\mu{\rm G}~~~& {\rm or}~\gamma(T_9=1) < 2.4\times 10^{-3}~~~& ({\rm D};~{\rm best}),
\label{eq1}
\end{eqnarray}

If one conservatively takes the constraint on $^4$He abundance by AOS10, the observation of D abundance provides the strongest upper limits on PMF.  The conservative upper limit from the mean value of QSO D/H ratio~\cite{Pettini:2012ph} is $B(0) < 1.5~\mu$G [$\gamma(T_9=1) < 4.6\times 10^{-3}$], while that from the best D/H measurement~\cite{Pettini:2012ph} is $B(0) < 0.78~\mu$G [$\gamma(T_9=1) < 2.4\times 10^{-3}$].  The latter limit is nearly identical to the previous estimation (corresponding to $\gamma(T_9=1) < 2.3\times 10^{-3}$ which is read from eq. (4) in Ref.~\cite{Grasso:1996kk}), while the former is less stringent than the former by a factor of two.  Previous constraints~\cite{Grasso:1996kk,Kernan:1995bz,Cheng:1996yi,Grasso:2000wj} have been derived neglecting changes in evolution of baryon-to-photon ratio $\eta$.  In our work, this effect is consistently taken into account, and the final $\eta$ value is fixed to the WMAP estimation.  The present result is, therefore, most precise.  Other improvements are updates of nuclear reaction rates, observational constraints on primordial abundances, and baryon-to-photon ratio.

\section{Discussion}\label{sec4}
The constraint derived in this study is related to the local field amplitude $B$ contributed from all wavelengths, and is not for that measured at any particular scale~\cite{Kernan:1995bz}.  The present amplitude of cosmological averaged field, i.e, $B_{\rm rms}$, (or the energy density of magnetic field) is defined~\cite{Kernan:1995bz} by
\begin{equation}
B(z)_{\rm rms}^2 =\frac{1}{V_{\rm H}}~\int_{V_{\rm H}}~d^3x|\bfB(\bfx,z)|^2,
\end{equation}
where
$z$ is the redshift,
$V_{\rm H}$ is the Hubble volume, and
$\bfx$ is the position vector.
The conservative constraint is then $B_{\rm rms}(0) <1.5~\mu$G.

Magnetic fields on some scales depend on the spatial structure of field.  The root mean square (rms) amplitude on scale $L$~\cite{Grasso:2000wj} is
\begin{equation}
B(L,z)_{\rm rms} =B(0)_{\rm rms} \left(1+z\right)^2 \left(\frac{L_0}{L}\right)^p,
\end{equation}
where
$L_0$ is the comoving coherence length, and
$p$ is a parameter determined from statistical properties of the magnetic field~\cite{Grasso:2000wj}.

The present constraint can be compared with those from other measurements summarized in Refs~\cite{Neronov:2009gh,Neronov:1900zz}.  We note that the new constraint [Eq. (\ref{eq1})] can be the strongest for small correlation scales of $L_0\lesssim 10^{-2}$~pc.  Direct constraints on magnetic field strength at smallest scales are derived from observations of Zeeman effect of HI, OH and CN in molecular clouds and HI diffuse clouds~\cite{2010ApJ...725..466C}.  The smallest upper limit on the radial component of magnetic field is $B_\parallel=0.0\pm0.9$~$\mu$G for an HI cloud seen in absorption against radio source 3C 348 (a usable data in Ref.~\cite{Heiles2004}).  Heiles and Troland used data of Zeeman-splitting of the 21 cm line~\cite{Heiles2004}, and estimated a median total field strength $B=6.0\pm 1.8$~$\mu$G for HI clouds with scales of typically $\mathcal{O}(0.01$ -- $10~{\rm pc})$, taking account of probability distribution function of total field strength $B$ and a random orientation of fields with respect to the line of sight~\cite{Heiles:2005xi}.  

The constraint on PMF from BBN studies cannot be directly compared with those from CMBR studies (e.g.~\cite{Yamazaki:2004vq,Yamazaki:2006bq,Yamazaki:2008gr,Yamazaki:2010nf}) , i.e., $B(1~{\rm Mpc}, 0)_{\rm rms}=0.85\pm 1.25$ nG~\cite{Yamazaki:2010nf}, since the CMBR limits are imposed on magnetic fields on scales larger than the horizon in the BBN epoch.~\cite{Grasso:2000wj}.  For example, when we adopt $L_0=100$~pc (the comoving Hubble horizon in the BBN epoch) and $p=3/2$ (which is derived in the assumption that a field vector performs a random walk in three dimensional space by steps of the physical scale $L_0$~\cite{1983PhRvL..51.1488H}), Eq. (\ref{eq1}) leads to
\begin{equation}
B(1~{\rm Mpc},0)_{\rm rms} < 1.5~{\rm pG}.
\end{equation}

Comparisons between constraints for different coherent lengths thus generally depend on statistical properties of magnetic field.  See Ref.~\cite{Widrow:2011hs} for a recent review for creation mechanisms of extragalactic magnetic fields and their problems.

\section{Conclusions}\label{sec5}
A new upper limit on the amplitude of primordial magnetic field (PMF) is derived by a comparison between a numerical calculation of elemental abundances in big bang nucleosynthesis and the latest constraints on abundances inferred from observations.  The newest nuclear reaction rates are adopted (Sec.~\ref{sec2}).  In addition, effects of PMF on the abundances are consistently taken into account in the numerical calculation with a formulation of physical variables in a magnetic field (Appendix \ref{appendix}).

We find that the existence of PMF increases abundances of $^3$He and $^6$Li, and decreases that of $^7$Li in the calculation for the baryon-to-photon ratio determined from the measurement of cosmic microwave background radiation with the Wilkinson Microwave Anisotropy Probe.  As a result of the rigorous calculation, we derive a constraint on the present amplitude of PMF, i.e., $B(0)<1.5~\mu$G [corresponding to the amplitude less than $4.6\times 10^{-3}$ times the critical magnetic field strength for electron at temperature $T=10^9$~K].

\appendix

\section{Formulae for effects of magnetic field on nucleosynthesis}\label{appendix}
\begin{widetext}
\begin{enumerate}
\item energy density\label{appendix1}

The energy density of magnetic field is
\begin{eqnarray}
\rho_B=\frac{B^2}{8\pi}=\frac{B_{\rm C}^2}{8\pi}\gamma^2,
 \label{eqa11}
\end{eqnarray}
where
$B$ is the amplitude of magnetic field, and
$\gamma=B/B_{\rm C}$ is the $B$ value in units of critical magnetic field, i.e., $B_{\rm C}=m_e^2/e=4.41\times 10^{13}$~G with $e$ the electric charge, and $m_e$ the electron mass.
This energy contributes to the total energy contents of the universe related to the Hubble expansion rate.  In this study, it is assumed that the primordial magnetic field (PMF) just attenuates by the cosmic expansion.  

\item thermodynamic variables of electron in a magnetic field\label{appendix2}

The number density, energy density and the pressure \footnote{The equation (2) of Ref.~\cite{Grasso:1996kk} was right, while the equation (2.15) of Ref.~\cite{Cheng:1996yi} was likely wrong.} of the electron and positron are given~\cite{Grasso:2000wj}, respectively, as 
\begin{eqnarray}
n_e(B)=\frac{eB}{(2\pi)^2} \sum_{n=0}^\infty (2-\delta_{n0}) \int_{-\infty}^{\infty} f_{\rm FD} (T_e,E_n)~dp_z,
 \label{eqa1}
\end{eqnarray}
\begin{eqnarray}
\rho_e(B)=\frac{eB}{(2\pi)^2} \sum_{n=0}^\infty (2-\delta_{n0}) \int_{-\infty}^{\infty} E_n~f_{\rm FD} (T_e,E_n)~dp_z,
 \label{eqa2}
\end{eqnarray}
\begin{eqnarray}
P_e(B)=\frac{eB}{(2\pi)^2} \sum_{n=0}^\infty (2-\delta_{n0}) \int_{-\infty}^{\infty} \frac{E_n^2-m_e^2}{3E_n}~f_{\rm FD} (T_e,E_n)~dp_z,
 \label{eqa3}
\end{eqnarray}
where
\begin{eqnarray}
f_{\rm FD} (T_e,E_n)=\frac{1}{1+\exp[\left(E_n(p_z)\mp \mu\right)/T_e]}
 \label{eqa4}
\end{eqnarray}
is the Fermi-Dirac distribution function at electron temperature $T_e$, and $E_n=[p_z^2+eB(2n+1+s)+m_e^2]^{1/2}$ is the energy of electron in the presence of a uniform field which is much smaller than the critical strength $B_{\rm C}$ \cite{Kernan:1995bz}, $n=0,~1,~...,~\infty$ and $s=\pm1$ are the principal and magnetic quantum numbers of the Landau level, respectively, and $\mu$ is the chemical potential of electron.  It has been assumed that the direction of field is the $z$-axis.

The above quantities can be rewritten in the form of
\begin{eqnarray}
n_e=\frac{m_e^3 \gamma}{2\pi^2} \sum_{n_{\rm S}=0}^\infty (2-\delta_{n_{\rm S}0}) \int_0^\infty~dk~\frac{1}{1+\mathrm{e}^{\epsilon z_e\mp \phi_e}},
 \label{eqa5}
\end{eqnarray}
\begin{eqnarray}
\rho_e=\frac{m_e^4 \gamma}{2\pi^2} \sum_{n_{\rm S}=0}^\infty (2-\delta_{n_{\rm S}0}) \int_0^\infty~dk~\epsilon~\frac{1}{1+\mathrm{e}^{\epsilon z_e \mp \phi_e}},
 \label{eqa6}
\end{eqnarray}
\begin{eqnarray}
P_e=\frac{m_e^4 \gamma}{2\pi^2} \sum_{n_{\rm S}=0}^\infty (2-\delta_{n_{\rm S}0}) \int_0^\infty~dk~\frac{k^2+2\gamma n_{\rm S}}{3\epsilon}~\frac{1}{1+\mathrm{e}^{\epsilon z_e \mp \phi_e}},
 \label{eqa7}
\end{eqnarray}
where
$k=p_z/m_e$, $\epsilon=(k^2+1+2\gamma n_{\rm S})^{1/2}$, $z_e=m_e/T_e$, and $\phi_e=\mu/T_e$ were defined.

Using the Euler-McLaurin formula \footnote{Kernan et al.~\cite{Kernan:1995bz} concluded that weak interaction rates decrease with increasing magnetic fields, while Cheng et al.~\cite{Cheng:1996yi} concluded that the rates increase with the fields.  This contradiction was caused since the former authors used the Euler-McLaurin expansion, while the latter used the Taylor expansions of the rates which are ill defined~\cite{Kernan:1996ab}.}, the number density and the energy density of electron and positron are given by
\begin{eqnarray}
n_e&=&\frac{T_e^3}{\pi^2} \left\{\int_0^\infty \frac{\rho^2~d\rho}{1+\exp(\sqrt{\mathstrut \rho^2+m_e^2/T_e^2} \mp \phi_e)} \right.\nonumber\\
&&\left.+ \frac{\gamma^2}{24}\left(\frac{m_e}{T_e}\right)^4 \int_0^\infty \frac{d\eta}{\sqrt{\mathstrut \eta^2+m_e^2/T_e^2}} \frac{1}{1+\cosh(\sqrt{\mathstrut \eta^2+m_e^2/T_e^2} \mp \phi_e)} \right.\nonumber\\
&&\hspace{2.em}\left. +O\left[\left(\gamma\frac{m_e^2}{T_e^2}\right)^4\right]\right\},
 \label{eqa8}
\end{eqnarray}
\begin{eqnarray}
\rho_e&=&\frac{T_e^4}{\pi^2} \left\{\int_0^\infty \frac{\rho^2 \sqrt{\mathstrut \rho^2+m_e^2/T_e^2}~d\rho}{1+\exp(\sqrt{\mathstrut \rho^2+m_e^2/T_e^2} \mp \phi_e)} \right.\nonumber\\
&&\hspace{2.em}+ \frac{\gamma^2}{24}\left(\frac{m_e}{T_e}\right)^4 \int_0^\infty d\eta \left[\frac{1}{1+\cosh(\sqrt{\mathstrut \eta^2+m_e^2/T_e^2} \mp \phi_e)}-\frac{2/\sqrt{\mathstrut \eta^2+m_e^2/T_e^2}}{1+\exp(\sqrt{\mathstrut \eta^2+m_e^2/T_e^2} \mp \phi_e)} \right] \nonumber\\
&&\hspace{2.em}\left. +O\left[\left(\gamma\frac{m_e^2}{T_e^2}\right)^4\right]\right\}.
 \label{eqa9}
\end{eqnarray}
These equations are the same as those derived in Ref.~\cite{Kernan:1995bz} except that ours are generalized versions including the electron chemical potential.  In order to follow in numerical calculations precisely the electron chemical potential, which becomes large at late time of BBN, it is kept in our formulation.  Adopting the Euler-McLaurin formula to the pressure of electron and positron, one can obtain the equation, i.e., 
\begin{eqnarray}
P_e&=&\frac{T_e^4}{3\pi^2} \left\{\int_0^\infty \frac{\rho^4~d\rho}{\sqrt{\mathstrut \rho^2+m_e^2/T_e^2}\left[1+\exp(\sqrt{\mathstrut \rho^2+m_e^2/T_e^2} \mp \phi_e)\right]} \right.\nonumber\\
&&\hspace{2.em}+ \frac{\gamma^2}{24}\left(\frac{m_e}{T_e}\right)^4 \int_0^\infty d\eta \left[\frac{\eta^2}{\eta^2+m_e^2/T_e^2}\frac{1}{1+\cosh(\sqrt{\mathstrut \eta^2+m_e^2/T_e^2} \mp \phi_e)}\right.\nonumber\\
&&\hspace{12.em}\left.-\frac{2/\sqrt{\mathstrut \eta^2+m_e^2/T_e^2}}{1+\exp(\sqrt{\mathstrut \eta^2+m_e^2/T_e^2} \mp \phi_e)}\left(2-\frac{\eta^2}{\eta^2+m_e^2/T_e^2}\right)\right] \nonumber\\
&&\hspace{2.em}\left. +O\left[\left(\gamma\frac{m_e^2}{T_e^2}\right)^4\right]\right\}.
 \label{eqa10}
\end{eqnarray}
Eqs. (\ref{eqa8}), (\ref{eqa9}) and (\ref{eqa10}) reproduce values for the case of no magnetic field when $B=0$ is input.

Time evolutions of following three variables as perturbations induced by a magnetic field are calculated.  The first is related to the asymmetry in number abundances of electron and positron, i.e,
\begin{eqnarray}
\frac{\pi^2}{2}\left[\frac{\hbar c}{m_e c^2}\right]^3 z^3 \Delta (n_{e^-}-n_{e^+})=\frac{1}{48}\int_0^\infty~d\eta f_n(\eta),
 \label{eqa12}
\end{eqnarray}
where
$\hbar$ is the Planck's constant,
$c$ is the light speed,
and
\begin{eqnarray}
f_n(\eta)\equiv \gamma^2 z^4 \frac{1}{\theta} \left[\frac{1}{1+\cosh(\theta-\phi_e)} -\frac{1}{1+\cosh(\theta+\phi_e)}\right]
 \label{eqa13}
\end{eqnarray}
and the parameter $\theta(\eta)\equiv \sqrt[]{\mathstrut \eta^2+m_e^2/T_e^2}$
was defined.  Partial derivatives of this function with respect to $T_9=T_e/(10^9~{\rm K})$, the neutrino temperature, i.e, $T_\nu$, and $\phi_e$ are given by
\begin{eqnarray}
\frac{\partial f_n(\eta)}{\partial T_9}&=&-\gamma^2\left\{
\frac{z^4}{\theta T_9}\left(4-\frac{z^2}{\theta^2}\right)
\left[\frac{1}{1+\cosh(\theta-\phi_e)} -\frac{1}{1+\cosh(\theta+\phi_e)}\right]
\right.\nonumber\\
&&\hspace{2.em}\left.+\frac{z^6}{\theta^2 T_9}\left[-\frac{\sinh(\theta-\phi_e)}{\left(1+\cosh(\theta-\phi_e)\right)^2} +\frac{\sinh(\theta+\phi_e)}{\left(1+\cosh(\theta+\phi_e)\right)^2}\right]\right\},
 \label{eqa14}
\end{eqnarray}
\begin{eqnarray}
\frac{\partial f_n(\eta)}{\partial T_\nu}=\frac{4\gamma^2 z^4}{\theta T_\nu}\left[\frac{1}{1+\cosh(\theta-\phi_e)} -\frac{1}{1+\cosh(\theta+\phi_e)}\right],
 \label{eqa15}
\end{eqnarray}
\begin{eqnarray}
\frac{\partial f_n(\eta)}{\partial \phi_e}=\gamma^2 z^4 \frac{1}{\theta} \left[\frac{\sinh(\theta-\phi_e)}{\left(1+\cosh(\theta-\phi_e)\right)^2} +\frac{\sinh(\theta+\phi_e)}{\left(1+\cosh(\theta+\phi_e)\right)^2}\right],
 \label{eqa16}
\end{eqnarray}
where
we used $\partial \gamma/\partial T_\nu= 2\gamma/T_\nu$.

The second variable is a perturbation in the total energy density of electron and positron induced by $B\neq 0$, i.e.,
\begin{eqnarray}
\Delta (\rho_{e^-}+\rho_{e^+})=\frac{m_e^4}{12\pi^2}\int_0^\infty~d\eta f_\rho(\eta),
 \label{eqa17}
\end{eqnarray}
where
\begin{eqnarray}
f_\rho(\eta)\equiv \gamma^2 \left\{\frac{1}{2\left(1+\cosh(\theta+\phi_e)\right)} +\frac{1}{2\left(1+\cosh(\theta-\phi_e)\right)} -\frac{1}{\theta}\left(\frac{1}{1+\mathrm{e}^{\theta+\phi_e}}+\frac{1}{1+\mathrm{e}^{\theta-\phi_e}}\right)\right\}\nonumber\\
 \label{eqa18}
\end{eqnarray}
was defined.  Partial derivatives of this function with respective to $T_9$, $T_\nu$ and $\phi_e$ are given by
\begin{eqnarray}
\frac{\partial f_\rho(\eta)}{\partial T_9}&=&-\frac{\gamma^2 z^2}{\theta T_9}\left\{
-\frac{\sinh(\theta+\phi_e)}{2\left(1+\cosh(\theta+\phi_e)\right)^2} -\frac{\sinh(\theta-\phi_e)}{2\left(1+\cosh(\theta-\phi_e)\right)^2} +\frac{1}{\theta^2} \left(\frac{1}{1+\mathrm{e}^{\theta+\phi_e}} +\frac{1}{1+\mathrm{e}^{\theta-\phi_e}}\right) \right.\nonumber\\
&&\hspace{2.em}\left. -\frac{1}{\theta}\left[-\frac{\mathrm{e}^{\theta+\phi_e}}{\left(1+\mathrm{e}^{\theta+\phi_e}\right)^2} -\frac{\mathrm{e}^{\theta-\phi_e}}{\left(1+\mathrm{e}^{\theta-\phi_e}\right)^2}\right] \right\},
 \label{eqa19}
\end{eqnarray}
\begin{eqnarray}
\frac{\partial f_\rho(\eta)}{\partial T_\nu}=\frac{4\gamma^2}{T_\nu}\left[\frac{1}{2\left(1+\cosh(\theta+\phi_e)\right)} +\frac{1}{2\left(1+\cosh(\theta-\phi_e)\right)} -\frac{1}{\theta}\left(\frac{1}{1+\mathrm{e}^{\theta+\phi_e}}+\frac{1}{1+\mathrm{e}^{\theta-\phi_e}}\right)\right],\nonumber\\
 \label{eqa20}
\end{eqnarray}
\begin{eqnarray}
\frac{\partial f_\rho(\eta)}{\partial \phi_e}&=&\gamma^2 \left\{-\frac{\sinh(\theta+\phi_e)}{2\left(1+\cosh(\theta+\phi_e)\right)^2} +\frac{\sinh(\theta-\phi_e)}{2\left(1+\cosh(\theta-\phi_e)\right)^2} \right.\nonumber\\
&&\left.-\frac{1}{\theta}\left[-\frac{\mathrm{e}^{\theta+\phi_e}}{\left(1+\mathrm{e}^{\theta+\phi_e}\right)^2} +\frac{\mathrm{e}^{\theta-\phi_e}}{\left(1+\mathrm{e}^{\theta-\phi_e}\right)^2}\right]\right\}.
 \label{eqa21}
\end{eqnarray}

The third variable is a perturbation in the total pressure of electron and positron, i.e.,
\begin{eqnarray}
\Delta (P_{e^-}+P_{e^+})=\frac{m_e^4}{36\pi^2}\int_0^\infty~d\eta f_P(\eta),
 \label{eqa22}
\end{eqnarray}
where
\begin{eqnarray}
f_P(\eta)&\equiv& \gamma^2 \left\{\frac{\eta^2}{\theta^2}\left[\frac{1}{2\left(1+\cosh(\theta+\phi_e)\right)} +\frac{1}{2\left(1+\cosh(\theta-\phi_e)\right)}\right] \right.\nonumber\\
&&\left.-\frac{2-\eta^2/\theta^2}{\theta}\left(\frac{1}{1+\mathrm{e}^{\theta+\phi_e}}+\frac{1}{1+\mathrm{e}^{\theta-\phi_e}}\right)\right\}
 \label{eqa23}
\end{eqnarray}
was defined.

\item density--temperature relation\label{appendix3}

In the BBN code~\cite{Kawano1992}, derivatives of $\phi_e$ with respect to $T_9$, $r=r(T_\nu)=\log(a^3)$ with $a$ the scale factor of the universe, and $S=\sum_i Z_i Y_i$ with $Z_i$ the charge and $Y_i$ the number ratio of nuclide $i$ to total baryon, respectively, are calculated and used.  In the calculation, we use the following equation for charge conservation in the universe:
\begin{eqnarray}
\frac{\pi^2}{2}\left[\frac{\hbar c}{m_e c^2}\right]^3 z^3 \left[n_{e^-}(B)-n_{e^+}(B)\right]=\frac{\pi^2}{2}\left[N_{\rm A} \left(\frac{\hbar c}{k}\right)^3 h S\right],
 \label{eqa24}
\end{eqnarray}
where
$n_{e^\mp}(B)=n_{e^\mp}(0)+\Delta n_{e^\mp}(B)$ is the number density of $e^\mp$ in an environment of magnetic field $B$,
$N_{\rm A}=6.02\times 10^{23}$ is the Avogadro's number, and
$k$ is the Boltzmann's constant.

The left and right-hand sides are denoted as $N=N(T_9, T_\nu, \phi_e)$ and $M=M(T_9, r, S)$, respectively.  Taking derivatives of both sides with respect to $T_9$, $r$ and $S$, three partial derivatives are obtained as in the case of no magnetic field~\cite{Kawano1992}:
\begin{eqnarray}
\left.\frac{\partial M}{\partial T_9}\right|_{r, S}= \frac{\partial N}{\partial T_9} + \frac{\partial N}{\partial \phi_e}\frac{\partial \phi_e}{\partial T_9} & \hspace{2.em} \Longrightarrow & \hspace{2.em} \frac{\partial \phi_e}{\partial T_9}=\left(\frac{\partial N}{\partial \phi_e}\right)^{-1} \left( \left.\frac{\partial M}{\partial T_9}\right|_{r, S} - \frac{\partial N}{\partial T_9} \right),\\
\left.\frac{\partial M}{\partial r}\right|_{T_9, S}= \frac{\partial N}{\partial T_\nu}\frac{\partial T_\nu}{\partial r} + \frac{\partial N}{\partial \phi_e}\frac{\partial \phi_e}{\partial r} & \hspace{2.em} \Longrightarrow & \hspace{2.em} \frac{\partial \phi_e}{\partial r}=\left(\frac{\partial N}{\partial \phi_e}\right)^{-1} \left( \left.\frac{\partial M}{\partial r}\right|_{T_9, S} +\frac{T_\nu}{3} \frac{\partial N}{\partial T_\nu} \right),\nonumber\\
\\
\left.\frac{\partial M}{\partial S}\right|_{T_9, r}= \frac{\partial N}{\partial \phi_e}\frac{\partial \phi_e}{\partial S} & \hspace{2.em} \Longrightarrow & \hspace{2.em} \frac{\partial \phi_e}{\partial S}=\left(\frac{\partial N}{\partial \phi_e}\right)^{-1} \left.\frac{\partial M}{\partial S}\right|_{T_9, r},
 \label{eqa252627}
\end{eqnarray}
where
we used $\partial T_\nu/\partial r= -T_\nu/3$.  The above derivatives are estimated utilizing Eqs. (\ref{eqa12}--\ref{eqa16}), and used in estimation of time evolution of the chemical potential parameter $\phi_e$.

$d\rho_e/d T_9$ is given by
\begin{eqnarray}
\frac{d \rho_e}{d T_9}=\frac{\partial \rho_e}{\partial T_9} + \frac{\partial \rho_e}{\partial \phi_e} \frac{d \phi_e}{d T_9} +\frac{\partial \rho_e}{\partial T_\nu}\frac{d T_\nu}{d T_9},
 \label{eqa28}
\end{eqnarray}
where $dT_\nu/dT_9$ can be described as
\begin{eqnarray}
\frac{d T_\nu}{d T_9}=-\frac{T_\nu}{3} \frac{dr}{dT_9}.
 \label{eqa29}
\end{eqnarray}

The conservation of energy for mixed matter of $\gamma$, $\nu$'s, $e^\pm$ and baryons leads~\cite{Kawano1992} to
\begin{eqnarray}
\frac{d r}{d T_9}=-\frac{\frac{d \rho_\gamma}{d T_9} + \left(\frac{\partial \rho_e}{\partial T_9} + \frac{\partial \rho_e}{\partial \phi_e} \frac{d \phi_e}{d T_9} \right) + \frac{d \rho_{\rm b}}{d T_9}}{\rho_\gamma + \frac{P_\gamma}{c^2} +\rho_e + \frac{P_e}{c^2} + \frac{P_{\rm b}}{c^2} +\frac{1}{dr/dt}\left(\left.\frac{d\rho_{\rm b}}{dt}\right|_{T_9} + \left.\frac{d\rho_e}{dt}\right|_{T_9} \right) -\frac{T_\nu}{3} \frac{\partial \rho_e}{\partial T_\nu}},
 \label{eqa30}
\end{eqnarray}
where
$\rho_\gamma$ and $\rho_{\rm b}$ are energy densities of photon and baryons, respectively, and
$P_\gamma$ and $P_{\rm b}$ are pressures of photon and baryons, respectively.  

Time derivative of the electron and positron energy density is given by
\begin{eqnarray}
\frac{1}{dr/dt} \left.\frac{d \rho_e}{dt} \right|_{T_9}=\frac{\partial \rho_e}{\partial \phi_e} \left(\frac{\partial \phi_e}{\partial r} + \frac{\partial \phi_e}{\partial S} \frac{\partial S}{\partial t} \frac{1}{dr/dt}\right) +\frac{\partial \rho_e}{\partial T_\nu} \left(-\frac{T_\nu}{3}\right).
 \label{eqa31}
\end{eqnarray}
Using Eqs. (\ref{eqa30}) and (\ref{eqa31}), we obtain
\begin{eqnarray}
\frac{d r}{d T_9}=-\frac{\frac{d \rho_\gamma}{d T_9} + \left(\frac{\partial \rho_e}{\partial T_9} + \frac{\partial \rho_e}{\partial \phi_e} \frac{d \phi_e}{d T_9} \right) + \frac{d \rho_{\rm b}}{d T_9}}{\rho_\gamma + \frac{P_\gamma}{c^2} +\rho_e + \frac{P_e}{c^2} + \frac{P_{\rm b}}{c^2} +\frac{1}{dr/dt}\left.\frac{d\rho_{\rm b}}{dt}\right|_{T_9} + \frac{\partial \rho_e}{\partial \phi_e} \left(\frac{\partial \phi_e}{\partial r} + \frac{\partial \phi_e}{\partial S} \frac{\partial S}{\partial t} \frac{1}{dr/dt}\right) -\frac{2}{3} T_\nu\frac{\partial \rho_e}{\partial T_\nu}}.\nonumber\\
 \label{eqa32}
\end{eqnarray}
The derivatives of $\rho_e$ [cf. Eqs. (\ref{eqa17}--\ref{eqa21})] and pressure value $P_e$ [Eq. (\ref{eqa22}--\ref{eqa23})] are input in this equation.  

It is clear that a PMF enhances energy density of $e^+$+$e^-$ [Eqs. (\ref{eqa17}) and (\ref{eqa18})].  The PMF, however, does not work on $e^\pm$, so that the enhanced energy of $e^\pm$ realizes by an energy transfer through interaction with thermal bath.  When the temperature decreases, the energy gain of $e^\pm$ decreases because of the weakening of PMF.  Resultingly, this loss of energy gain heats the thermal bath.  This effect is taken into account in Eq. (\ref{eqa32}) \footnote{Kernan, Starkman and Vachaspati~\cite{Kernan:1995bz} have suggested an error of Cheng et al.~\cite{Cheng:1993kz} that the PMF is included in the work-energy equation for photon.}.

\item initial value of electron chemical potential\label{appendix4}

Using Eqs. (\ref{eqa12}), (\ref{eqa24}) and the fact of $S\cong Y_{\rm p}$ in hot environments before the nucleosynthesis, we obtain an equation for initial value of electron chemical potential, i.e.,
\begin{eqnarray}
\phi_e\sim \frac{\pi^2}{2}\left[N_{\rm A} \left(\frac{\hbar c}{k}\right)^3 h Y_{\rm p}\right] \frac{1}{z^3} \left[\sum_n (-1)^{n+1} n L(nz) + \frac{z\gamma^2}{24} \int_0^\infty~d\eta~\frac{1}{\theta}~\frac{\sinh \theta}{(1+\cosh \theta)^2}\right]^{-1},\nonumber\\
 \label{eqa33}
\end{eqnarray}
where
$L(z)=K_2(z)/z$~\cite{Kawano1992} was defined with $K_2(z)$ the modified Bessel functions.

\item note

Some transformations in equations are used in order to avoid appearances of divergences in a numerical calculation.  They include $1/[1+\cosh(\theta-\phi_e)]-1/[1+\cosh(\theta+\phi_e)]=[2\sinh \phi_e/(1+\cosh(\theta-\phi_e)][(1-\mathrm{e}^{-2\theta})/(\mathrm{e}^{\phi_e}+2\mathrm{e}^{-\theta}+\mathrm{e}^{-2\theta-\phi_e})]$, $\sinh x/(1+\cosh x)^2=2(1-\mathrm{e}^{-x})/[\mathrm{e}^x (1+\mathrm{e}^{-x})^3]$ and $\mathrm{e}^x/(1+\mathrm{e}^x)^2=1/[\mathrm{e}^x(1+\mathrm{e}^{-x})^2]$.

\end{enumerate}
\end{widetext}

\section{Effect on nuclear reaction rates}\label{appendix_2}
It was claimed that relatively weak magnetic fields reduce collision rates of nuclear reactions by a factor of two through alignments of directions in which charged nuclei can move~\cite{1986PPCF...28..857I,1987PPCF...29..951I}.  The claim is, however, wrong~\cite{1987PPCF...29..949J,1987PPCF...29..953H} since it was based on a completely-mistaken treatment of 2D space perpendicular to the field direction, and cases of $B=0$ and $B\neq 0$ are not treated consistently.

Precise reaction rates are derived below.  The point is that a target is hit by projectiles coming from all directions although the distribution function of projectile charged particles are quantized.

\subsection{Nuclear distribution function}
Magnetic fields would affect nuclear reaction rates through a discretization of momentum on the plane perpendicular to the field direction ($z$-axis).  The Zeeman splitting also realizes in magnetic fields.  It is, however, neglected in this study since situations of large fields are eventually excluded from light element abundances deviated through their effects on the cosmic expansion rate.  Energy levels of charged nuclides are then given by
\begin{equation}
 E_n=\left[p_z^2+ZeB(2n+1)+m^2\right]^{1/2}\sim m+\frac{1}{2m}\left[p_z^2+ZeB(2n+1)\right],
\label{eqb1}
\end{equation}
where
$m$ and $Z$ are the mass and the charge number of a nuclide, respectively.

The nuclear distribution function is discretized similarly to the case of electron as
\begin{equation}
 \int \frac{g}{(2\pi)^3}d^3p~{\mathrm e}^{-(E-\mu_{\rm C})/T}\longrightarrow ZeB \sum_{n=0}^\infty \int \frac{g}{(2\pi)^2}dp_z~{\mathrm e}^{-(E_n-\mu_{\rm C})/T},
\label{eqb2}
\end{equation}
where
$g$ is the statistical weight, and
$\mu_{\rm C}$ is the chemical potential of the nuclide.  The number density of nuclide $i$ in a Landau level $n_{{\rm L}i}$ and in a momentum range between $p_{z i}$ and $p_{z i}+dp_{z i}$ is given by
\begin{equation}
 n_i(n_{{\rm L}i},p_{z i})dp_{z i}=\frac{g_i}{(2\pi)^2}Z_ieB \exp\left[-\frac{\mu_{{\rm C} i}-m_i-Z_ieB(2n_{{\rm L}i}+1)/(2m_i)}{T}\right] \exp\left(-\frac{p_{z i}^2}{2m_iT}\right)dp_{z i}.
\label{eqb3}
\end{equation}
The total number density is given by
\begin{eqnarray}
 n_i&=&\frac{u_ig_i}{(2\pi)^2} Z_ieB \exp\left[-\frac{\mu_{{\rm C} i}-m_i-Z_ieB/(2m_i)}{T}\right]\int_{-\infty}^\infty \exp\left(-\frac{p_{z i}^2}{2m_iT}\right)dp_{z i}\nonumber\\
&=&\frac{g_i u_i m_i^{1/2}T^{1/2}}{2^{3/2}\pi^{3/2}}Z_ieB \exp\left[-\frac{\mu_{{\rm C} i}-m_i-Z_ieB/(2m_i)}{T}\right],
\label{eqb4}
\end{eqnarray}
where 
\begin{equation}
u_i\equiv \sum_{n_{{\rm L} i}=0}^\infty \exp\left(-\frac{n_{{\rm L} i}Z_i eB}{m_iT}\right)=\left[1-\exp\left(-\frac{Z_i eB}{m_i T}\right)\right]^{-1}
\label{eqb5}
\end{equation}
was defined.

The fraction in number density is derived from Eqs. (\ref{eqb4}) and (\ref{eqb5}) as
\begin{equation}
 \frac{n_i(n_{{\rm L}i},p_{z i})dp_{z i}}{n_i}=\frac{1}{2^{1/2}\pi^{1/2}m_i^{1/2}T^{1/2}}u_i^{-1} \exp\left(-\frac{p_{z i}^2}{2m_iT}-\frac{n_{{\rm L}i}Z_i eB}{m_i T}\right)dp_{z i}.
\label{eqb6}
\end{equation}

\subsection{Rates of reactions between charged particles}
The thermal average of reaction rate is described as
\begin{equation}
 \langle \sigma v \rangle=\frac{1}{n_1 n_2}~\sum_{n_{{\rm L} 1}=0}^\infty~\sum_{n_{{\rm L} 2}=0}^\infty \int n_1(n_{{\rm L} 1},p_{z1})dp_{z1}~n_2(n_{{\rm L} 2},p_{z2})dp_{z2}~\sigma(E) v \frac{d\cos\theta}{2},
\label{eqb7}
\end{equation}
where
$v$ is the relative velocity of nuclides 1 and 2,
$E$ is the kinetic energy in the center of mass (CM) system, and
$\theta$ is the angle between momentum vectors of nuclides 1 and 2 on the plane perpendicular to the magnetic field.

The velocity vectors of nuclides $i$ is described as $\bfv_i=(\bfv_{\perp i},~v_{z i})$.  The angle is then given by $\cos\theta=\bfv_{\perp 1}\cdot \bfv_{\perp 2}/(v_{\perp 1} v_{\perp 2})$ with amplitudes of the vectors $v_{\perp i}=|\bfv_{\perp i}|$.  The relative velocity is given by $v=[v_{\perp 1}^2 -2v_{\perp 1} v_{\perp 2}\mu +v_{\perp 2}^2 +(v_{z1}-v_{z2})^2]^{1/2}$ with $\mu=\cos\theta$.  The CM kinetic energy is $E=\mu_{\rm red} v^2/2$ with the reduced mass $\mu_{\rm red}=m_1 m_2/(m_1+m_2)$.

Substituting Eq. (\ref{eqb6}) in Eq. (\ref{eqb7}), we obtain an equation,
\begin{eqnarray}
 \langle \sigma v \rangle&=&\frac{1}{2^2\pi m_1^{1/2} m_2^{1/2}T} {u_1}^{-1}{u_2}^{-1}\sum_{n_{{\rm L} 1}=0}^\infty \sum_{n_{{\rm L} 2}=0}^\infty \exp\left[-\left(\frac{n_{{\rm L} 1}Z_1}{m_1}+\frac{n_{{\rm L} 2}Z_2}{m_2}\right)\frac{eB}{T}\right] \nonumber\\
&&\times \int_{-\infty}^\infty dp_{z1}~\int_{-\infty}^\infty dp_{z2}~\int_{-1}^1 d\mu~\exp\left(-\frac{m_1v_{z1}^2+m_2v_{z2}^2}{2T}\right)~\sigma(E) v.
\label{eqb8}
\end{eqnarray}
Momentum variables, i.e., $p_{z 1}$ and $p_{z 2}$, are transformed to the CM momentum and the relative momentum.  Integration over the CM momentum is performed in the equation, and we obtain,
\begin{eqnarray}
 \langle \sigma v \rangle&=&\frac{1}{2^{1/2} \pi^{1/2}}\frac{\mu_{\rm red}^{1/2}}{T^{1/2}} \left[1-\exp\left(-\frac{Z_1eB}{m_1T}\right)\right] \left[1-\exp\left(-\frac{Z_2eB}{m_2T}\right)\right] \nonumber\\
&&\times \sum_{n_{{\rm L} 1}=0}^\infty\sum_{n_{{\rm L} 2}=0}^\infty \exp\left[-\left(\frac{n_{{\rm L} 1}Z_1}{m_1}+\frac{n_{{\rm L} 2}Z_2}{m_2}\right)\frac{eB}{T}\right]\int_0^\infty dv_{z{\rm r}}~\int_{-1}^1 d\mu~\exp\left(-\frac{\mu_{\rm red}v_{z{\rm r}}^2}{2T}\right)~\sigma(E) v,\nonumber\\
\label{eqb9}
\end{eqnarray}
where
$v_{z{\rm r}}\equiv v_{z1}-v_{z2}$ is the relative velocity in the direction of the field.

Velocities on the plane perpendicular to the field are discretized as $v_{\perp i}^2=(2n_{{\rm L} i}+1)Z_ieB/m_i^2$.  The relative velocity can then be described by
\begin{equation}
 v=\left\{eB\left[\frac{(2n_{{\rm L} 1}+1)Z_1}{m_1^2} + \frac{(2n_{{\rm L} 2}+1)Z_2}{m_2^2} -2\mu\frac{\sqrt[]{\mathstrut (2n_{{\rm L} 1}+1) Z_1~(2n_{{\rm L} 2}+1) Z_2}}{m_1 m_2}\right] +v_{z{\rm r}}^2\right\}^{1/2}.
\label{eqb10}
\end{equation}

\subsection{Rates of reactions between a neutron and charged particles}

In reactions of neutron, discrete momenta of only charged particles are taken into account.  The rate is described as
\begin{equation}
 \langle \sigma v \rangle=\frac{1}{n_1}~\sum_{n_{{\rm L} 1}=0}^\infty~\int n_1(n_{{\rm L} 1},p_{z1})dp_{z1} \int \left(\frac{1}{2\pi m_n T}\right)^{3/2}~d^3p~\exp\left(-\frac{p_n^2}{2m_nT}\right)~\sigma(E)v,
\label{eqb11}
\end{equation}

Variable transformations from $p_{z 1}$ and $p_{z n}$ to $p_{z {\rm G}}$ (CM momentum) and $p_{z {\rm r}}$ (relative momentum) are performed, and an integration over $p_{z {\rm G}}$ is computed.  The reaction rate is then rewritten to be
\begin{eqnarray}
 \langle \sigma v \rangle&=&\frac{1}{2^{5/2} \pi^{3/2} \mu_{\rm red}^{1/2} m_n T^{3/2}}~u_1^{-1}~\sum_{n_{{\rm L} 1}=0}^\infty \exp\left(-\frac{n_{{\rm L} 1} Z_1eB}{m_1T}\right)\nonumber\\
&&\times \int_{-\infty}^\infty dp_{z{\rm r}} \int_{-1}^1 d\mu \int_{-\infty}^\infty dp_{xn}~dp_{yn}~\exp\left(-\frac{p_{z{\rm r}}^2}{2\mu T}-\frac{p_{xn}^2+p_{yn}^2}{2m_nT}\right)~\sigma(E)v.
\label{eqb12}
\end{eqnarray}
We perform an integration over azimuth angle on the ($p_{xn},p_{yn}$) plane and trivial transformations from momenta to velocities, and obtain an expression, i.e.,
\begin{eqnarray}
 \langle \sigma v \rangle&=&\frac{1}{2^{1/2}\pi^{1/2}}\frac{\mu_{\rm red}^{1/2} m_n}{T^{3/2}} \left[1-\exp\left(-\frac{Z_1eB}{m_1T}\right)\right]~\sum_{n_{{\rm L} 1}=0}^\infty \exp\left(-\frac{n_{{\rm L} 1} Z_1eB}{m_1T}\right)\nonumber\\
&&\times \int_0^\infty dv_{z{\rm r}} \int_0^\infty dv_{\perp n} \int_{-1}^1 d\mu~v_{\perp n}~\exp\left(-\frac{\mu_{\rm red} v_{z {\rm r}}^2+m_n v_{\perp n}^2}{2T}\right)~\sigma(E)v,
\label{eqb13}
\end{eqnarray}
where
the relative velocity is given by 
\begin{equation}
 v=[v_{\perp 1}^2 + v_{\perp n}^2 -2\mu v_{\perp 1} v_{\perp n} + v_{z{\rm r}}^2]^{1/2}.
\label{eqb14}
\end{equation}
The kinetic energy in the CM system is given by $E=\mu_{\rm red} v^2/2$.

The sum in the reaction rate is transformed to an integration using the Euler-McLaurin formula.  The integration form of reaction rate is given by
\begin{eqnarray}
 \langle \sigma v \rangle&=&\frac{1}{2^{1/2}\pi^{1/2}}\frac{\mu_{\rm red}^{1/2} m_n}{T^{3/2}} \left[1-\exp\left(-\frac{Z_1eB}{m_1T}\right)\right]\nonumber\\
&&\times \left\{\int_0^\infty dn_{{\rm L} 1} \exp\left(-\frac{n_{{\rm L} 1} Z_1eB}{m_1T}\right) \int_0^\infty dv_{z{\rm r}} \exp\left(-\frac{\mu_{\rm red} v_{z {\rm r}}^2}{2T}\right) \right.\nonumber\\
&&\left.\hspace{2.em}\times \int_0^\infty dv_{\perp n}~v_{\perp n}\exp\left(-\frac{m_n v_{\perp n}^2}{2T}\right) \int_{-1}^1 d\mu~\sigma(E)v \right.\nonumber\\
&&\left. + \frac{1}{2} \left(1+\frac{Z_1eB}{6m_1T}\right) \int_0^\infty dv_{z{\rm r}} \exp\left(-\frac{\mu_{\rm red} v_{z {\rm r}}^2}{2T}\right) \right.\nonumber\\
&&\left.\hspace{2.em}\times \int_0^\infty dv_{\perp n}~v_{\perp n}\exp\left(-\frac{m_n v_{\perp n}^2}{2T}\right) \int_{-1}^1 d\mu~\left[\sigma(E)v\right]_{n_{{\rm L} 1}=0} \right.\nonumber\\
&&\left. -\frac{1}{12} \frac{Z_1eB}{m_1^2} \int_0^\infty dv_{z{\rm r}} \exp\left(-\frac{\mu_{\rm red} v_{z {\rm r}}^2}{2T}\right) \right.\nonumber\\
&&\left.\hspace{2.em}\times \int_0^\infty dv_{\perp n}~v_{\perp n}\exp\left(-\frac{m_n v_{\perp n}^2}{2T}\right) \int_{-1}^1 d\mu~\mu\left(1-\frac{v_{\perp n}}{v_{\perp 1}}\mu \right)\left[\frac{\partial \left[\sigma(E)v\right]}{\partial E}\right]_{n_{{\rm L} 1}=0} \right\}.~~~~~~~~~
\label{eqb15}
\end{eqnarray}
Two terms scaling as $eB$ are induced by magnetic field $B$, and they disappear in the limit of no field, i.e., $B=0$.  

\subsubsection{$^7$Be($n,p$)$^7$Li}

We check the reaction rate of $^7$Be($n,p$)$^7$Li for example.  Rates are calculated with Eq. (\ref{eqb13}).  The masses of neutron and $^7$Be nucleus is $m_n=0.939565$ GeV and $m_{^7{\rm Be}}=6.534184$ GeV~\cite{Audi2003}.  Although the cross section would be changed in magnetic fields through the momentum quantization, that effect is neglected and cross section values in no fields are taken from Ref.~\cite{Descouvemont2004} approximately.

\begin{figure}
\begin{center}
\includegraphics[width=8.0cm,clip]{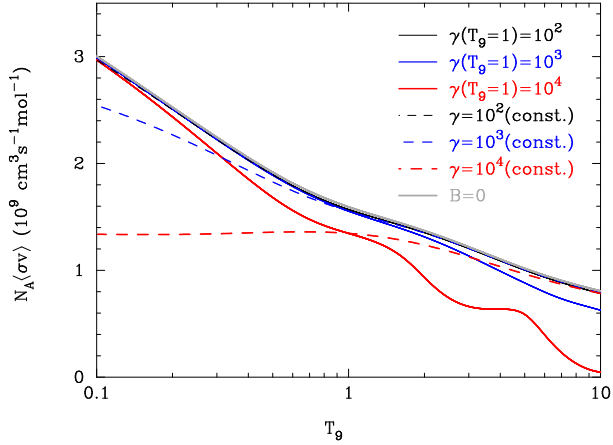}
\caption{Reaction rates of $^7$Be($n,p$)$^7$Li as a function of the temperature $T_9\equiv T/(10^9~{\rm K})$.  Solid dark lines correspond to cases of attenuating magnetic field for $B/(B_{\rm C}T_9^2)=10^2$, $10^3$ and $10^4$, in order of increasing line width, respectively.  Dashed lines correspond to cases of constant magnetic field for $B/B_{\rm C}=10^2$, $10^3$ and $10^4$, in order of increasing line width, respectively.  Solid gray line corresponds to the rate in no magnetic field. \label{fig4}}
\end{center}
\end{figure}

Figure~\ref{fig4} shows calculated rates of $^7$Be($n,p$)$^7$Li as a function of $T_9$.  Solid dark lines correspond to cases in which magnetic fields decrease with time because of the cosmic expansion as $B\propto T^2$.  Field amplitudes are  $B/(B_{\rm C}T_9^2)=10^2$, $10^3$ and $10^4$, in order of increasing line width (from the top to the bottom), respectively.  Dashed lines, on the other hand, correspond to cases of constant magnetic field of $B/B_{\rm C}=10^2$, $10^3$ and $10^4$, in order of increasing line width (from the top to the bottom), respectively.  Solid gray line corresponds to the rate in no magnetic field.  The uppermost solid and dashed lines are hardly distinguishable from the solid gray line.


In the attenuating magnetic field case (solid dark lines), the field effect is larger in higher temperatures.  The discretization effect is roughly determined from the index factor, i.e., $Z_1eB/(m_1T)$, in the exponential in Eq. (\ref{eqb13}).  The field amplitude ($\propto T^2$) decreases faster than the temperature in the universe ($\propto T^1$) does.  The index is, therefore, larger in higher temperature.  In the constant magnetic field case (dashed lines), the field effect is larger in lower temperatures conversely.

In somewhat high magnetic field, the minimum energy of nuclear Landau level is higher than the thermal energy of the universe, $\sim T$.  The average of rate then receives a contribution from large CM energies which originate from large relative velocities [Eq. (\ref{eqb14})].  Since the reaction rate at higher energies is roughly smaller as for the reaction $^7$Be($n,p$)$^7$Li, the existence of field decreases the reaction rate.  The lowest solid dark line in Fig. \ref{fig4} has two bumps at $1\lesssim T_9\lesssim 2$ and $4\lesssim T_9\lesssim 7$.  These bumps correspond to peaks in reaction rates produced by the $3^+$ resonant states of $^9$Be$^\ast$ at resonance energies $E_{\rm r}=0.33$ MeV and $2.66$ MeV~\cite{Adahchour2003}.  At $T_9=1.75$ and $T_9=5.2$, the minimum CM energies of the ground Landau level, i.e., $E_{\rm min}=\mu_{\rm red}Z_1eB/(2m_1^2)$ [cf. Eq. (\ref{eqb14})], are $0.3$ MeV and $2.7$ MeV, respectively.

As observed above, magnetic fields can affect thermonuclear reaction rates.  Large amplitudes of magnetic fields as assumed in Fig. \ref{fig4} are, however, excluded from incredibly fast expansion of universe (see Fig. \ref{fig2}).  The effect of magnetic field on nuclear reaction rates can thus be neglected.

\begin{acknowledgments}
This work is supported by Grant-in-Aid for Scientific Research from the
Ministry of Education, Science, Sports, and Culture (MEXT), Japan,
No.22540267 and No.21111006 (Kawasaki) and JSPS Grant No.21.6817
 (Kusakabe) and by World Premier International Research Center
 Initiative (WPI Initiative), MEXT, Japan.
\end{acknowledgments}

\bibliography{reference}



\end{document}